\documentclass[prb,superscriptaddress,floats,floatfix]{revtex4}

\usepackage{graphicx}
\usepackage{gensymb}
\usepackage{amsmath}
\usepackage{bm}
\begin{document}

\title  {An accurate single-electron pump based on a highly tunable silicon quantum dot}

\author{Alessandro Rossi}\email{a.rossi@unsw.edu.au}
\affiliation{School of Electrical Engineering \& Telecommunications, The University of New South Wales, Sydney 2052, Australia}
\author{Tuomo Tanttu}
\author{Kuan Yen Tan}
\affiliation{QCD Labs, COMP Centre of Excellence, Department of Applied Physics, Aalto University, P.O. Box 13500, 00076 Aalto, Finland}
\author{Ilkka Iisakka}
\affiliation{Centre for Metrology and Accreditation (MIKES), P.O. Box 9, 02151 Espoo, Finland}
\author{Ruichen Zhao}
\author{Kok Wai Chan}
\affiliation{School of Electrical Engineering \& Telecommunications, The University of New South Wales, Sydney 2052, Australia}
%\altaffiliation{Present address: Graphene Research Centre, School of Physics, Faculty of Science, National University of Singapore}
\author{Giuseppe C. Tettamanzi}
\author{Sven Rogge}
\affiliation{School of Physics and Australian Research Council Centre of Excellence for Quantum Computation and Communication Technology, The University of New South Wales, Sydney 2052, Australia}
\author{Andrew S. Dzurak}
\thanks{These authors contributed equally to this work}
\affiliation{School of Electrical Engineering \& Telecommunications, The University of New South Wales, Sydney 2052, Australia}
\author{Mikko M\"{o}tt\"{o}nen}\thanks{These authors contributed equally to this work}
\affiliation{QCD Labs, COMP Centre of Excellence, Department of Applied Physics, Aalto University, P.O. Box 13500, 00076 Aalto, Finland}

\date{\today}

\begin{abstract}
Nanoscale single-electron pumps can be used to generate accurate currents, and can potentially serve to realize a new standard of electrical current based on elementary charge. Here, we use a silicon-based quantum dot with tunable tunnel barriers as an accurate source of quantized current. The charge transfer accuracy of our pump can be dramatically enhanced by controlling the electrostatic confinement of the dot using purposely engineered gate electrodes. Improvements in the operational robustness, as well as suppression of non-adiabatic transitions that reduce pumping accuracy, are achieved via small adjustments of the gate voltages. We can produce an output current in excess of 80 pA with experimentally determined relative uncertainty below 50 parts per million.
\end{abstract}

\maketitle

As early as one and a half centuries ago, J. C. Maxwell envisaged the need for a system of standards based on phenomena at the atomic scale and directly related to invariant constants of nature.~\cite{max} However, Maxwell could not anticipate that, in order to harness the behaviour of the world at the nanometer scale, a completely new physical interpretation was needed, namely, quantum mechanics. At first, the laws of quantum mechanics seemed to reveal fundamental limits to the accuracy of physical measurements. Concepts like the Heisenberg uncertainty principle, which imposes intrinsic fluctuations on the values of non-commuting observables, and the wavefunction collapse, responsible for the randomization of a system configuration after performing a measurement, appeared to be at odds with the requirement of deterministic consistency that is paramount for metrological purposes. Nevertheless, quantum-based systems are today acknowledged as the most stable and reliable metrological tools, as they can be strongly intertwined with fundamental constants. Exquisitely quantum-mechanical phenomena such as the ac Josephson effect~\cite{josephson} and the quantum Hall effect~\cite{hall} have paved the way towards new and more reliable reference standards for the units of voltage and resistance, respectively.\\\indent Major efforts are currently ongoing to re-define the unit of electrical current, the ampere (A), in terms of the elementary charge, $e$, by means of quantum technologies~\cite{e_metro,peko}. A practical implementation of this standard may be the electron pump, a device in which a quantum phenomenon, namely tunnelling, and classical Coulomb repulsion, are combined to control the transfer of an integer number of elementary charges. This device ideally generates a quantized output current, $I_\textup{P}=nef$, where $n$ is an integer and $f$ is the frequency of an external periodic drive. Several enabling technologies have already been developed including metal/oxide tunnel barrier devices~\cite{devo,zimmer}, normal-metal/superconductor turnstiles~\cite{pek,ville}, graphene double quantum dots~\cite{graphene}, donor-based pumps~\cite{lans,roche,giu}, silicon-based quantum dot pumps~\cite{ono,fuji04,fuji08,kokwai,xavier} and GaAs-based quantum dot pumps~\cite{kou,saw,blume,PTB_PRB,samPRB,PTB_apl,sam_par,giblin_njp,giblin_natcom}. To date, the latter scheme provides the lowest uncertainty of 1.2 parts per million (ppm) yielding current in excess of 150 pA~\cite{giblin_natcom}. This remarkable result has been attained by exploiting the beneficial effect of an external magnetic field on the quantization accuracy~\cite{samPRB,PTB_apl,giblin_natcom}. An effect that has been ascribed to the suppression of non-adiabatic excitations and the reduction of initialization errors, as a consequence of the increased magnetic confinement.~\cite{npl_prb, npl_prl}\\
\begin{figure*}
\includegraphics[scale=0.6]{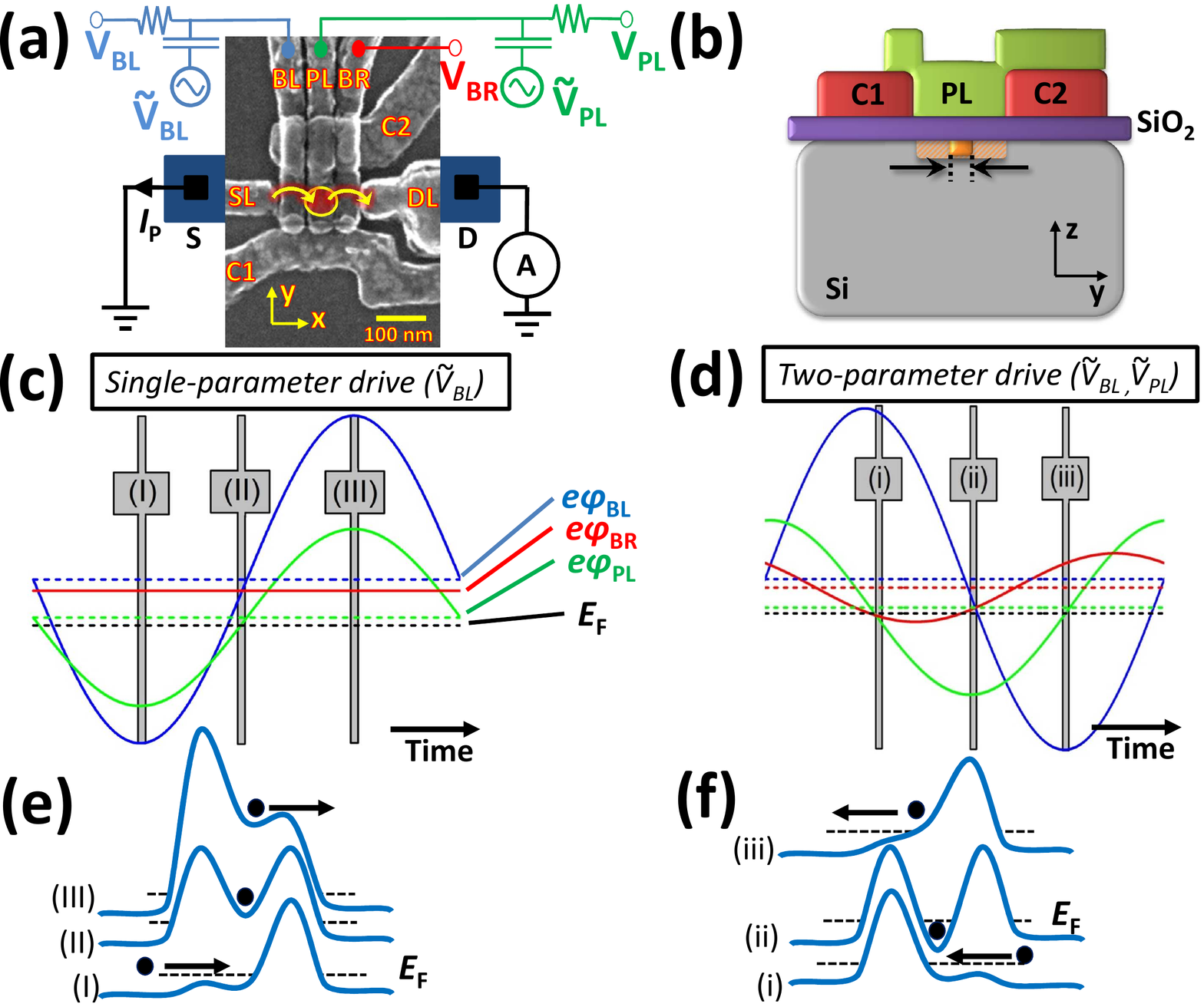}
\caption{(a) SEM image of an electron pump device similar to the one used in the experiments. The quantum dot is formed under gate PL in the proximity of the region highlighted in red/yellow. The yellow arrows show the direction of electron transport from/to reservoirs and through the dot. The blue pads indicate source (S) and drain (D) ohmic contacts. The electrical measurement set-up is also sketched. (b) Schematic cross-sectional view of the device at the dot position. Area in orange represents the planar extension of the dot. The confinement effect due to C1 and C2 is sketched. (c) Time dependence of the potentials (solid lines) at the right barrier ($\varphi_{\textup{BR}}$), left barrier ($\varphi_{\textup{BL}}$) and dot centre ($\varphi_{\textup{PL}}$) in the case of a single sinusoidal drive, $V_\textup{BL}+\widetilde{V}_\textup{BL} \sin(2\pi f t)$. Dashed lines indicate the dc components of the potentials and the Fermi level (black). (d) Similar diagrams as in (c) but for two sinusoidal drives, $V_\textup{BL}+\widetilde{V}_\textup{BL} \sin(2\pi f t)$ and $V_\textup{PL}+\widetilde{V}_\textup{PL} \sin(2\pi f t+\Delta\phi)$. Signals are arranged to exemplify a condition of pumping from drain to source. (e) Schematic energy diagrams illustrating the pumping mechanism during three stages of the cycle illustrated in (c) and showing transport of electrons from source to drain. (f) Similar diagrams as in (e) but for the two-parameter drive shown in (d), showing transport from drain to source.}
  \label{fgr:device}
\end{figure*}
\indent Here, we discuss a single-electron quantum dot pump that has allowed us to demonstrate an experimental uncertainty below 50 ppm for an output current as high as $80$~pA, without the need for magnetic confinement. The device investigated has been fabricated by means of metal-oxide-semiconductor (MOS) technology in silicon. The gate electrodes that are used to define the quantum dot have been purposely designed to achieve high tunability of the planar confinement. This degree of flexibility proves to be essential to improve the fidelity and robustness of the output current, as well as in mitigating the detrimental effects triggered by non-adiabatic transitions. In the future, simultaneous magnetic and electrostatic confinements as well as the use of arbitrarily defined waveforms~\cite{giblin_natcom} should further enhance the performance of this type of pump. From a practical viewpoint, an electrical current standard based on an industry-compatible silicon MOS process would benefit from readily available large-scale integration techniques to enable parallelization, and on-chip calibration, detection, and control circuits. Our pump, with its unprecedented low level of uncertainty for silicon devices, provides great potential for the implementation of a silicon-based current standard.
\\\indent The device employed in this work is realised on a substrate made of high-purity, near-intrinsic silicon which is thermally oxidized to grow a 8-nm-thick high-quality SiO$_2$ gate oxide. Three layers of aluminium gates are defined via electron-beam lithography to selectively accumulate electrons at the Si/SiO$_2$ interface~\cite{angus,lim}. Isolation between gate layers is achieved using thermally grown Al$_\textup{y}$O$_\textup{x}$~\cite{angus}. A scanning-electron-micrograph (SEM) image of a sample similar to the one used for the experiments is shown in Fig.~\ref{fgr:device}(a). A quantum dot is induced under the plunger gate, PL, by applying a positive bias voltage, while gates BL and BR define tunnel barriers upon application of lower voltages. Strong planar confinement for the dot potential is attained by independently controlling the voltages of two confinement gates C1 and C2, as illustrated in Fig.~\ref{fgr:device}(b). Two-dimensional electron gas (2DEG) reservoirs are formed under gates SL and DL; these extend to heavily $n$-doped regions defined in the substrate via phosphorus diffusion and act as source/drain ohmic contacts.  This design provides significant operational flexibility. For example, gates C1 and C2 are employed to control the confinement of the dot with minimal effect on the tunnel barrier potential profiles. Devices of this kind have already been proven to be very effective in precisely controlling the quantum dot energy level spectrum~\cite{our_natcom}. As we discuss below, important advantages in the context of charge pumping can also be obtained.\\
\indent In order to implement the pumping cycle, two configurations with either one or two sinusoidal driving voltages have been used, as schematically shown in Fig.~\ref{fgr:device}(c) and (d), respectively. Low-temperature bias tees have been employed to superimpose the ac excitations on the dc gate bias, as depicted in Fig.~\ref{fgr:device}(a). In the case of the one-signal drive, the ac excitation is applied to gate BL so that the potential profile at the left barrier, $\varphi_{\textup{BL}}$, is periodically modulated. However, due to capacitive cross coupling, the potential at the dot, $\varphi_{\textup{PL}}$, is also modulated. We assume the potential at the right barrier, $\varphi_{\textup{BR}}$, to be approximately unaffected. Thus, the potentials of both the left barrier and the dot contain isofrequential time-dependent components of nearly equal phase but different amplitudes. The energy diagrams in Fig.~\ref{fgr:device}(e) sketch how an electron enters the dot from the source (I), becomes trapped in the potential well (II), and finally is ejected to the drain (III). The yellow arrows in Fig.~\ref{fgr:device}(a) depict the electron flow for this configuration which produces positive quantized current given the conventions used in the measurement set-up.\\\indent In the case of the two-signal drive, two sinusoidal voltages at the same frequency, but typically different amplitudes and phases, are applied to gates BL and PL. Unlike in the previous situation, the potential at the right barrier will be modulated due to cross coupling with PL. As a result, all three potentials will have a time-dependent component at the same frequency but the effective phase differences and amplitudes are functions of the amplitudes and phases of the driving voltages (see Section A of the Supporting Material). By appropriately setting these parameters, it is possible to achieve pumping in either direction.\\
Figure~\ref{fgr:device}(f) schematically represents the periodic transfer of electrons from drain to source for amplitudes and phases of the two sinusoidal excitations consistent with the experimental findings. In Section A of the Supporting Information, we provide experimental data where bidirectional quantized current is achieved in this manner. The capability of reversing the direction of pumping at will is of paramount interest in the perspective of implementing a real-time error detection protocol which involves loading and unloading electrons onto and out of an island for charge sensing~\cite{zimmer,yama,fricke}. However, in this work, we mainly focus on using these two protocols to pump electrons from source to drain because it yields the highest accuracy in our case. We stress that neither of the pumping protocols used requires a drain--source bias, $V_\textup{DS}$. Hence $V_\textup{DS}=0$ in all the measurements reported, unless otherwise stated.\\
\indent All experiments were performed by cooling the sample in a self-made plastic dilution refrigerator with a base temperature of about 80 mK. However, during the experiments with high-frequency drive, the cryostat temperature typically increased by a few hundred mK, due to attenuation in the cables and connectors together with the limited cooling power of the cryostat. Importantly, the performance of the device appeared to be very good despite the elevated temperature. This robustness results from the large charging energy ($\approx 12$~meV) of the very small quantum dot configured here. In addition, the sample underwent several thermal cycles between cryogenic temperatures and room temperature.  After each cool-down, the experimental parameters were iteratively tuned to achieve the most accurate output current. We observed only minor deviations in the optimal bias conditions over tens of cool-downs. The sinusoidal excitations were generated using an arbitrary-waveform generator (Tektronix AWG7122B) which was synchronized to an external rubidium atomic clock (SRS FS725).\\\indent
The confinement gates can be utilized to enhance the planar electrostatic confinement of the quantum dot and hence the pumping accuracy. Figure ~\ref{fgr:fits}(a) shows the pumped current, $I_\textup{P}$, as a function of the plunger and C2 gate voltages obtained with a single sinusoidal drive at 100 MHz. A wide plateau region where the current is stable at $I_\textup{P}=ef$ indicates that one electron per cycle is accurately transferred from source to drain. The inset shows the current at the flattest point of the same plateau, given by the minimum of d$I_\textup{P}$/d$V_\textup{PL}$, as a function of the driving frequency. These data confirm the expected linear relationship between current and frequency, $I_\textup{P}=nef$. In Fig.~\ref{fgr:fits}(a) there is a notable widening of the plateau region with decreasing C2 gate voltage. In general, by decreasing the confinement gate voltages, one expects to reduce the size of the quantum dot due to modification of the potential profile. This hypothesis is confirmed by measuring the dot charging energy for different confinement gate voltages. As we show in Section B of the Supporting Information, reducing the confinement bias voltage by a few tens of mV can increase the charging energy by a few meV. This is consistent with our observation of robust charge quantization, since thermally activated errors can be largely suppressed by increasing the charging energy. Furthermore, tight confinement may lead to the suppression of initialization errors~\cite{npl_prb,casc}.\\\indent
In order to study more quantitatively the enhancement in pumping accuracy, we compare the experimental data with the decay cascade model~\cite{casc}. In Fig.~\ref{fgr:fits}(b), the measured average number of electrons transferred per cycle, $n=I_\textup{P}/ef$, is shown as a function of $V_\textup{PL}$ for different values of $V_\textup{C2}$. The plateaux become clearly wider and the transition current steeper with decreasing confinement gate voltage. By fitting the experimental data to the analytical result derived from the decay cascade model~\cite{casc}

\begin{figure*}
\includegraphics[scale=0.8]{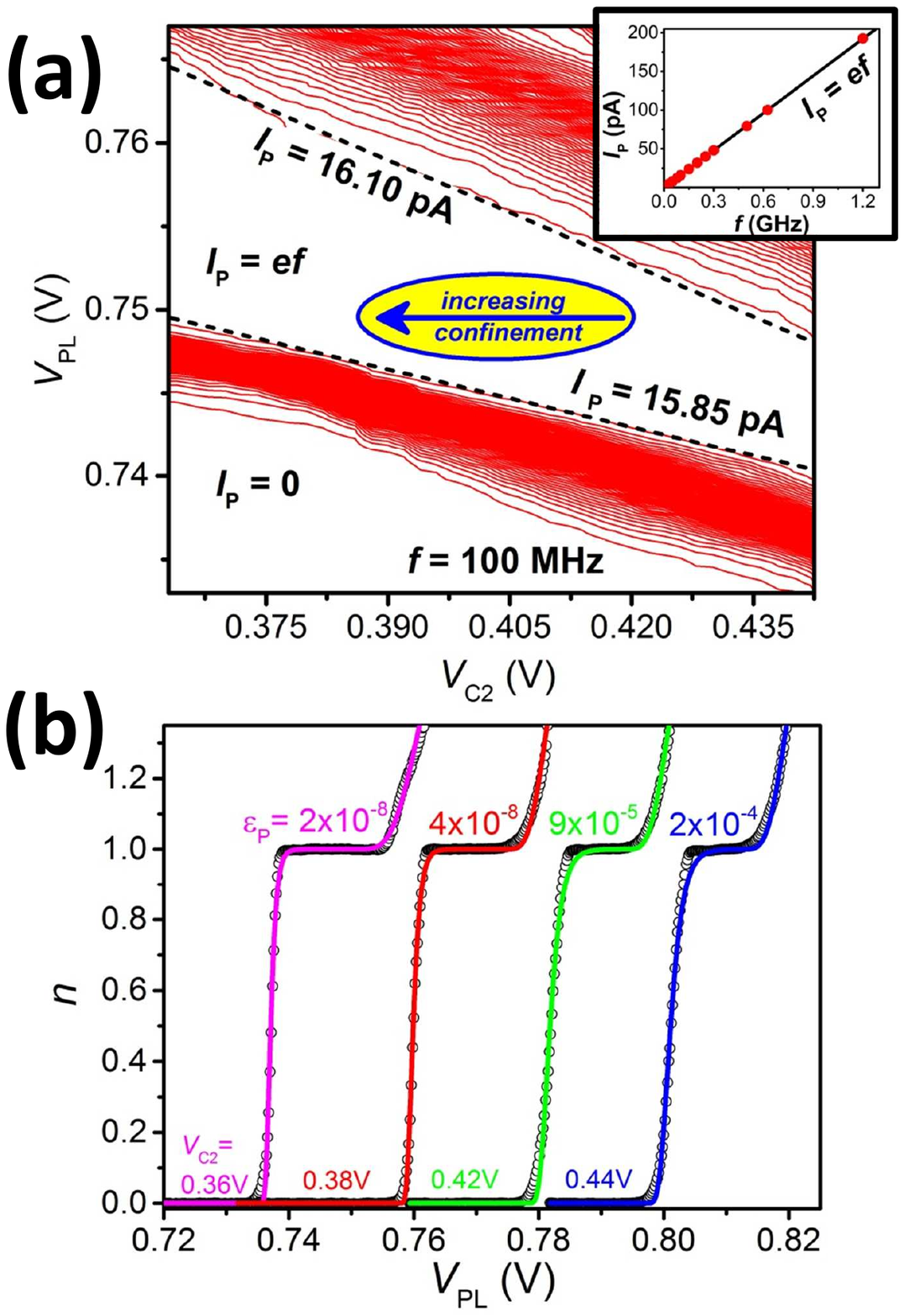}
\caption{(a) Contour plot of the pumped current as a function of the voltages of the plunger gate and C2 confinement gate for a single sinusoidal drive at $f=100$~MHz, $\widetilde{V}_\textup{BL}\approx 0.2$~V$_\textup{pp}$ and $V_\textup{C1}=-0.05$~V. Contour lines are in steps of 250 fA. Dashed lines are guides for the eye to roughly indicate the extremities of the plateaux. Inset: Frequency dependence of the pumped current at the first plateau. Individual data points are evaluated at the minima of d$I_\textup{P}$/d$V_\textup{PL}$. The solid line indicates the linear relationship expected for pumping one electron per cycle. (b) Measured average number of pumped electrons per cycle as a function of $V_\textup{PL}$ for different $V_\textup{C2}$ (open circles), and fits to the decay cascade model (solid lines). The traces are shifted horizontally for clarity.}
    \label{fgr:fits}
\end{figure*}

\begin{figure*}
\includegraphics[scale=0.9]{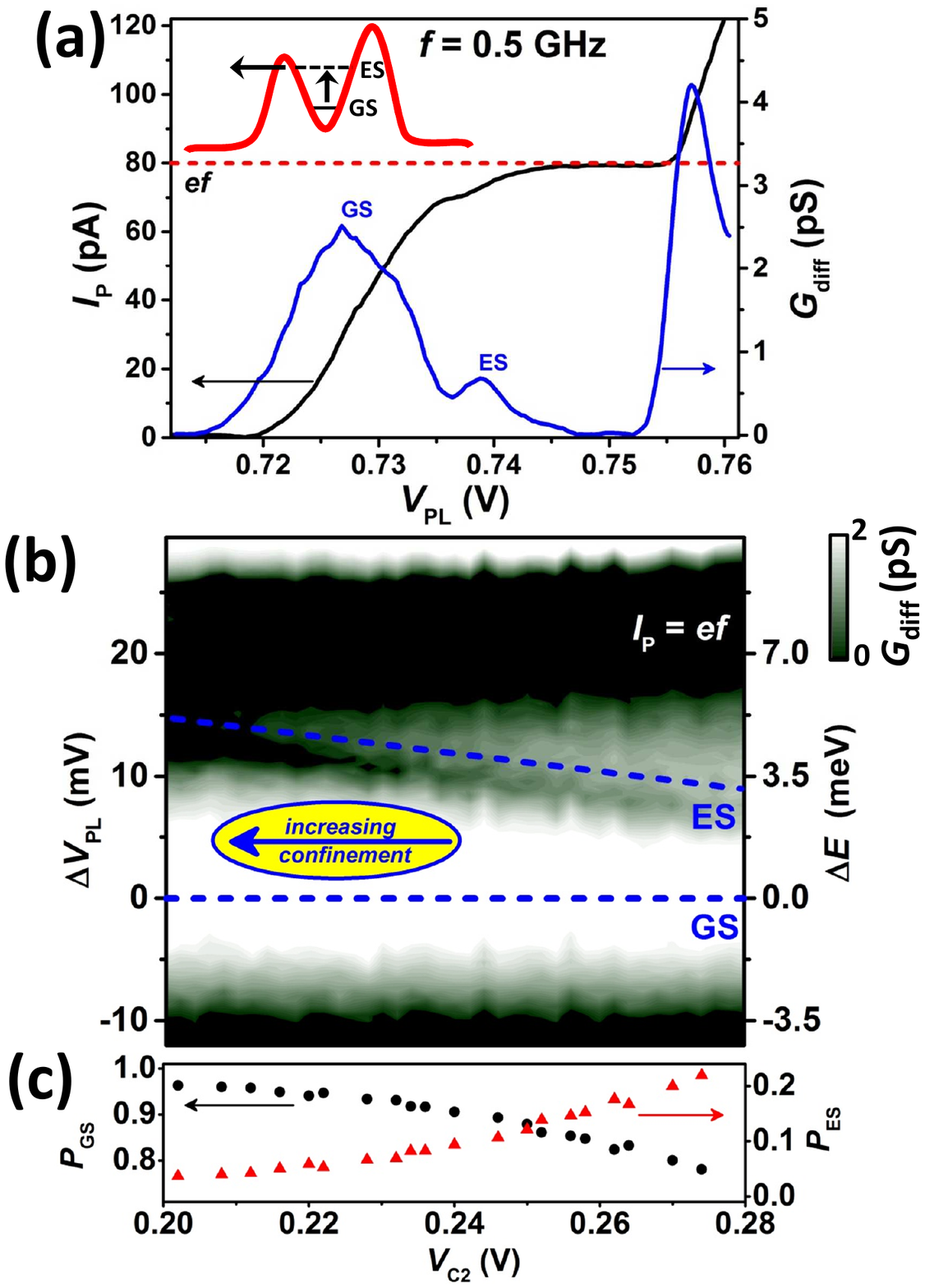}
  \caption{(a) Pumped current (black trace), $I_\textup{P}$, and differential trans-conductance (blue trace), $G_\textup{diff}\equiv$ d$I_\textup{P}$/d$V_\textup{PL}$, as functions of the plunger gate voltage for $V_\textup{C1}=-0.30$~V, $V_\textup{C2}=0.25$~V and two-parameter drive at $f=0.5$~GHz, $\Delta\phi=-20$ deg, and   $\widetilde{\textup{V}}_\textup{PL}\approx 0.014$~V$_\textup{pp}$, $\widetilde{\textup{V}}_\textup{BL}\approx 0.26$~V$_\textup{pp}$. Inset: Potential landscape illustrating a non-adiabatic excitation responsible for back-tunnelling events. (b) Colour map of $G_\textup{diff}$ as a function of   $V_\textup{C2}$ and plunger gate voltage referenced to the position of GS ($\Delta V_\textup{PL}=0$). Dashed lines are guides for the eye to highlight the relative position of the ground and excited states. (c) Occupation probabilities of ground (circles) and excited (triangles) states as a function of $V_\textup{C2}$. }
  \label{fgr:nonadiab}
\end{figure*}

\begin{equation}
  n_\textup{FIT}(V_\textup{PL})=\sum\limits_{i=1}^2\textup{exp(-exp(}-aV_\textup{PL}+\Delta_i)),
  \label{eqn:casc}
\end{equation}
where $a$ and $\Delta_i$ are different real-valued fitting parameters for each trace, one can extract the theoretical pumping error, $\epsilon_\textup{P}=1-n_\textup{FIT}(V^{\textup{*}}_\textup{PL})$, at the point $V^{\textup{*}}_\textup{PL}$ where d$n_\textup{FIT}$/d$V_\textup{PL}$ is minimized~\cite{giblin_natcom, casc}. Solid lines in Fig.~\ref{fgr:fits}(b) indicate the individual fitted traces accompanied by the theoretical pumping errors, which decrease by four orders of magnitude with $V_\textup{C2}$ decreasing by 80 mV. We have observed similar behaviour with respect to gate C1. This result is remarkable because it is achieved by solely controlling the electrostatic confinement of the quantum dot with very small bias adjustments. This is in stark contrast to GaAs pumps, for which magnetic fields as high as 14 T have been applied to have a comparable improvement~\cite{giblin_natcom}. Although the theoretical error for the most confined configuration is as low as $2\times 10^{-8}$, we do not claim this to be a faithful estimate of the performance of our pump. Discrepancies between this model and experimental observations have been reported~\cite{giblin_njp,giblin_natcom}. \\\indent
Another advantage arising from the tunability of the confinement potential is the possibility of mitigating the detrimental effects of single-particle excited states on charge quantization. In line with GaAs-based pumps~\cite{npl_prb, npl_prl}, the quality of the current plateaux deteriorates with increasing pumping frequency because of a partial reduction in current that  takes the form of an additional plateau at $I_\textup{P}<ef$. Figure~\ref{fgr:nonadiab}(a) shows the pumped current and the differential trans-conductance $G_\textup{diff}\equiv$~d$I_\textup{P}$/d$V_\textup{PL}$ at $f=0.5$~GHz for the case of the two driving signals discussed above. The relative amplitude and phase of the sine waves are selected to pump electrons from source to drain. Besides the wide current plateau at $I_\textup{P}=ef\approx 80.1$~pA, another narrower plateau at a smaller current value  $I_\textup{P}\approx 69$~pA is also visible. In the differentiated data, this feature appears as a small peak immediately following the large one produced by the main falling edge of the current. As discussed by Kataoka and coworkers~\cite{npl_prl}, the partial loss in quantization can be ascribed to non-adiabatic excitations from the electronic ground state to the first excited state [inset of Fig.~\ref{fgr:nonadiab}(a)] due to fast modulation of the confinement potential~\cite{math}. Promoted electrons have relatively high probability to tunnel back to the source rather than being correctly pumped to drain and, hence, they contribute to the overall transport cycle as a reduction of current.\\\indent In Fig.~\ref{fgr:nonadiab}(b), we show the differential trans-conductance at $f=0.5$~GHz as a function of $V_\textup{PL}$ and of the confinement gate bias $V_\textup{C2}$. The $V_\textup{PL}$ axis is shifted such that the position of the ground state establishes the reference for the whole dataset. The measurements indicate that the separation between the ground and excited state monotonically increases as the confinement becomes more pronounced, as one would expect from a simple quantum mechanical model of a particle in a potential well. The effective single-particle level separation, $\Delta E\equiv E_\textup{ES}-E_\textup{GS}$, can be evaluated for different confinement gate voltages as $\Delta E(V_\textup{C2})=\Delta V_\textup{PL}(V_\textup{C2})\times \alpha_\textup{PL}$, where $\Delta V_\textup{PL}$ is the voltage difference of the occurrences of the ES and GS peaks, and $\alpha_\textup{PL}\approx 0.35$~eV/V is the plunger lever arm as extracted from Coulomb diamond measurements. As shown in Fig.~\ref{fgr:nonadiab}(b), the energy separation varies from $\Delta E\approx 3.1$~meV at $V_\textup{C2}=0.280$V to $\Delta E\approx 5.1$~meV at $V_\textup{C2}=0.205$V. As expected, the probability of an electron to occupy an excited state decreases as the energy gap increases, and hence non-adiabatic excitations as well as the consequent accuracy-disruptive back tunnelling events become less likely. The occupation probabilities of the ground ($P_\textup{GS}$) and excited ($P_\textup{ES}$) states at the ES plateaux can be extracted from the values of the current at the GS and ES plateaux~\cite{npl_prl}, denoted by $I_\textup{GS}$ and $I_\textup{ES}$, respectively. We calculate $P_\textup{ES}\equiv 1-P_\textup{GS}=(I_\textup{ES}-I_\textup{GS})/I_\textup{GS}$. Figure~\ref{fgr:nonadiab}(c) shows that the excited state approaches zero occupation probability, as the confinement becomes more prominent. This is clear evidence of the suppression of non-adiabatic transitions. The importance of this result lies in the fact that a configurable electrostatic confinement may result in a significant aid towards fast pumping, in combination with the use of arbitrarily-shaped waveforms~\cite{giblin_natcom}.
\begin{figure*}
\includegraphics[scale=0.65]{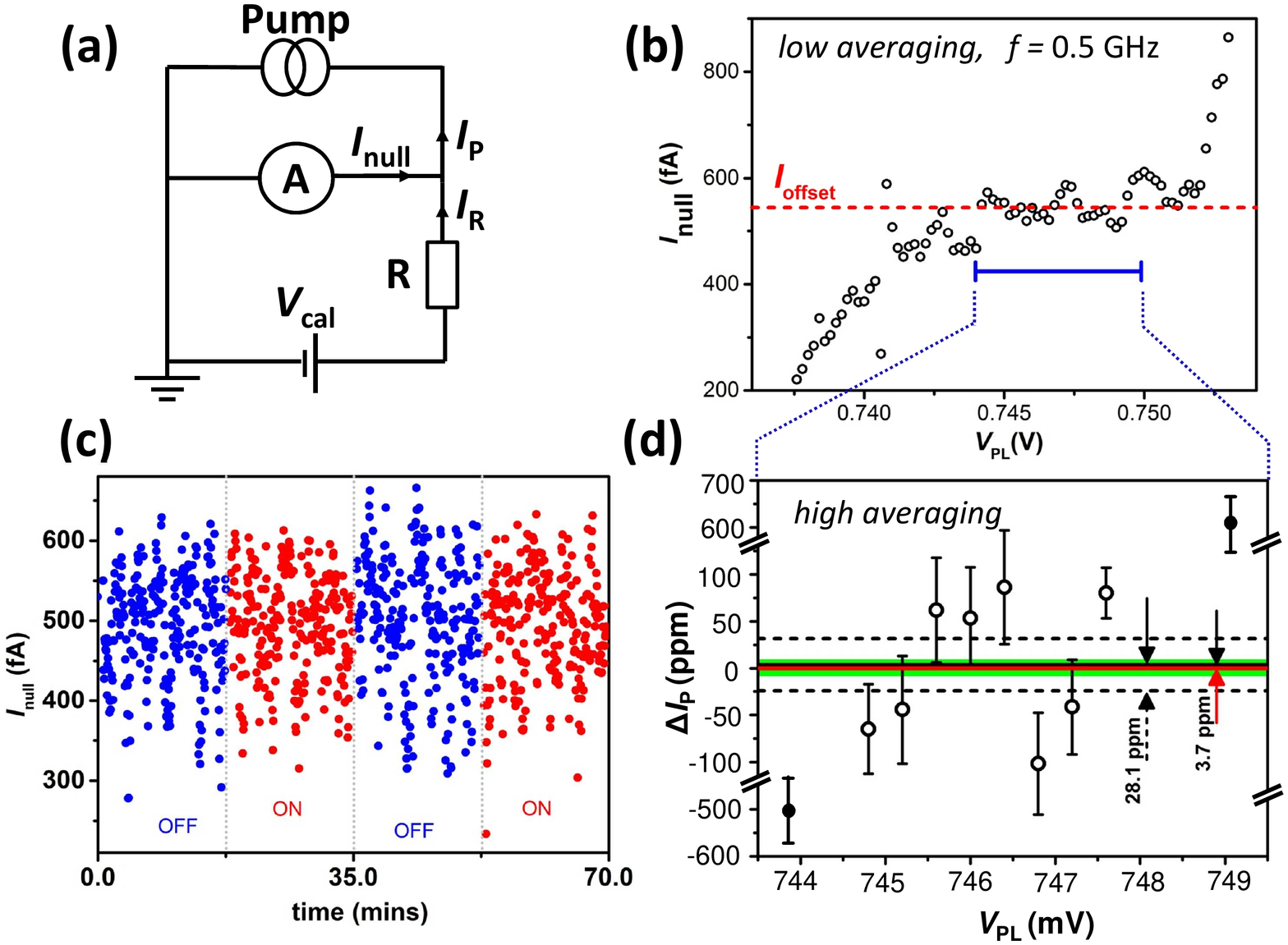}
%\centering
\caption{(a) Electrical circuit used for high-accuracy measurements. The pump device is indicated as a current source. The output of a direct voltage calibrator ($V_\textup{cal}$) is applied to a temperature-controlled resistor ($R=1.000037$~G$\Omega$) to produce a reference current, $I_\textup{R}$, that balances the pump current, $I_\textup{P}$. (b) Coarse measurement of the null current, $I_\textup{null}$, as a function of the plunger gate voltage. The red dashed line is a guide for the eye to indicate the offset current in the measurement circuitry. The blue bar indicates the plateau range over which high-accuracy measurements in (d) are carried out. (c) Measured null current for two consecutive ON-OFF cycles at $V_\textup{PL}=0.746$~V. During the OFF half-cycle (blue circles) both the voltage calibrator and the wave generator are switched off, and only the offset current is detected by the ammeter. During the ON half-cycle (red circles), both instruments are operational and the ammeter measures both the offset and the deviation of the pumped current from $V_\textup{cal}/R$. (d) High-accuracy measurements of the relative pumping error as a function of $V_\textup{PL}$. Circles (error bars) represent the mean (1$\sigma$ random uncertainty) of readings distributed over 2 ON-OFF cycles. The first $\approx 60$~s are discarded to remove possible transients in the circuitry. The black solid (dashed) line indicates the mean (error of the mean) of the distribution for the eight points on the plateau (open circles). The red solid line at zero is the reference for error-free pumping. The green area indicates the $\pm 6.3$~ppm systematic uncertainty of the measurement system. For all panels two-parameter sinusoidal drive is used with $f=0.5$~GHz, $\Delta\phi=-32$ deg,  $\widetilde{\textup{V}}_\textup{PL}\approx 0.015$~V$_\textup{pp}$, $\widetilde{\textup{V}}_\textup{BL}\approx 0.28$~V$_\textup{pp}$, $V_\textup{C1}=-0.36$~V and $V_\textup{C2}=0.19$~V. }
  \label{fgr:accu}
\end{figure*}
\\\indent Next, we discuss high-accuracy measurements which provide convincing evidence of quantized electron transport with a relative uncertainty below 50 ppm. As discussed above, our quantum dot pump performs at its best when under pronounced electrostatic confinement. Hence, we bias the confinement gates at  $V_\textup{C1}=-0.36$~V and $V_\textup{C2}=0.19$~V. In order to measure pumped currents with high accuracy, the measurement set-up is re-arranged as depicted in Fig.~\ref{fgr:accu}(a). The current produced by the pump is combined at a null detector (Femto DDPCA-300) with a reference current, $I_\textup{R}$, of opposite polarity generated by a calibrated voltage, $V_\textup{cal}$, across a temperature-controlled  $1~\textup{G}\Omega$ resistor, $R$ (Measurements International 4310HR). Given that $V_\textup{cal}$ is set to give $I_\textup{R}\equiv V_\textup{cal} / R=ef $, the very small current flowing through the null detector, $I_\textup{null}\equiv I_\textup{P}-I_\textup{R}$, is a direct evaluation of the deviation of the pumped current from the expected value. Most importantly, by implementing this null-detection configuration, variations in the measured current as a consequence of environmental effects, such as drifts in the gain of the null detector with temperature, have a negligible effect on the measurements. Therefore, the main contributions to the total systematic uncertainty, $u_\textup{S}$, of our measurements are those relevant to the calibration of $R$  and $V_\textup{cal}$ denoted as $u_\textup{S,$\Omega$}$ and $u_\textup{S,V}$ respectively. The resistor and voltage source have been calibrated at the Centre for Metrology and Accreditation of Finland (MIKES) with traceability to the quantum Hall resistance standard and Josephson voltage standard. For the resistor, we have $R=1.000037$~$\textup{G}\Omega$ with 1$\sigma$ relative uncertainty $u_\textup{S,$\Omega$}=5.7$~ppm. The reference voltage is obtained as the output of a direct voltage calibrator (Fluke 5440B) with relative uncertainty $u_\textup{S,V}=2.5$ ppm at $V_\textup{cal}=80.1088$~mV, which is used to generate the reference current in our experiments. This leads to the evaluation of the total systematic uncertainty of our measurement system as
\begin{equation}
  u_\textup{S}\equiv\sqrt{u^2_\textup{S,$\Omega$}+u^2_\textup{S,V}}=6.3~\textup{ppm.}
  \label{eqn:systematic}
\end{equation}
\\\indent Figure~\ref{fgr:accu}(b) shows a low-accuracy measurement of $I_\textup{null}$ for two driving voltages at 0.5 GHz. It is evident that the current at the plateau is not vanishing due to an offset in the null measurement circuitry of about 550 fA. To account for this artefact, both $I_\textup{P}$ and $I_\textup{R}$ are simultaneously switched on and off~\cite{giblin_natcom} and repeated readings of $I_\textup{null}$ are taken in cycles, as illustrated in Fig.~\ref{fgr:accu}(c). The offset current is eliminated by calculating the average null current as $\overline{I}_\textup{null}=\overline{I}_\textup{ON}-\overline{I}_\textup{OFF}$, where $\overline{I}_\textup{ON(OFF)}$ is the mean of the distribution of the readings during the ON (OFF) half-cycles. From high-accuracy measurements of $I_\textup{null}$ in a reduced range of $V_\textup{PL}$ with respect to Fig.~\ref{fgr:accu}(b), the relative pumping error 
\begin{equation}
\Delta I_\textup{P}\equiv \frac{I_\textup{P}-ef}{ef}\equiv \frac{ \overline{I}_\textup{null}+V_\textup{cal}/R-ef}{ef}, 
   \label{eqn:error}
\end{equation}
can be directly extracted, as shown in Fig.~\ref{fgr:accu}(d). Each data point has been averaged for 70 minutes over 2 ON-OFF cycles and the error bars represent $\pm 1\sigma$ random uncertainty calculated as the error of the mean of the values obtained from the two cycles. By neglecting data points that do not lie at the current plateau, the mean and the error of the mean of the remaining eight points is calculated. The total 1$\sigma$ relative uncertainty of our measurement can then be quoted as $29$~ppm, where the dominant component is of random type arising from the error of the mean of the eight data points. This indicates that improvements to the noise floor of our set-up are required; preliminary tests indicate that most of this excess noise is likely originating from the cabling between the null-detector and the 1 G$\Omega$ resistor. Although high-accuracy measurements to verify the invariance of the pumped current as a function of all the experimental parameters are beyond the scope of this work, a low-accuracy investigation has revealed good quantization over a broad range of all the adjustable variables (see Section C of the Supporting Material).\\
\indent In summary, we have demonstrated single-electron pumping through a silicon quantum dot with sinusoidal excitations. By controlling the electrostatic confinement of the dot using gate electrodes, the output current robustness is enhanced, and the loss of quantization due to non-adiabatic excitations can be greatly suppressed. By exploiting this high flexibility of our design, we demonstrate at 0.5 GHz experimental uncertainty  below 50 ppm, which is unprecedented in silicon-based charge pumps. Future use of magnetic confinement and arbitrarily-shaped driving voltages, as well as improvements to our measurement platform are likely to produce further increases in both the speed of electron transfer and the pumping accuracy, promoting this kind of electron pump as an ideal candidate for the realization of a quantum current standard. Finally, we note that the ability to accurately transport electrons at high speed is a crucial requirement for spin-based quantum computing architectures~\cite{holle}. Silicon has emerged as a very promising system material due to recent demonstrations of coherent spin manipulation~\cite{pla,maune}. In this context, silicon-based charge pumping could also enable coherent transport of quantum information within a quantum computer architecture.

\section*{ACKNOWLEDGMENTS}

We acknowledge useful discussions with F. Hudson, M. Veldhorst, Y. Sun, A. Manninen, and A. Kemppinen.
We acknowledge financial support from the Australian Research Council (Grant No. DP120104710), the Academy of Finland (Grant No. 251748, 135794, 272806) and support from the Australian National Fabrication Facility for device fabrication. We acknowledge the provision of facilities and technical support by Aalto University at Micronova Nanofabrication Centre. A.R. acknowledges support from the Australian Nanotechnology Network for a travel grant. G. C. T. acknowledges the Australian Research Council DECRA scheme (Grant No. DE120100702). S.R. acknowledges the Australian Research Council FT scheme (Grant No. FT100100589).

\section*{SUPPLEMENTARY INFORMATION}
\subsection{Bidirectional pumping}
\begin{figure*}[h]
\includegraphics[scale=0.75]{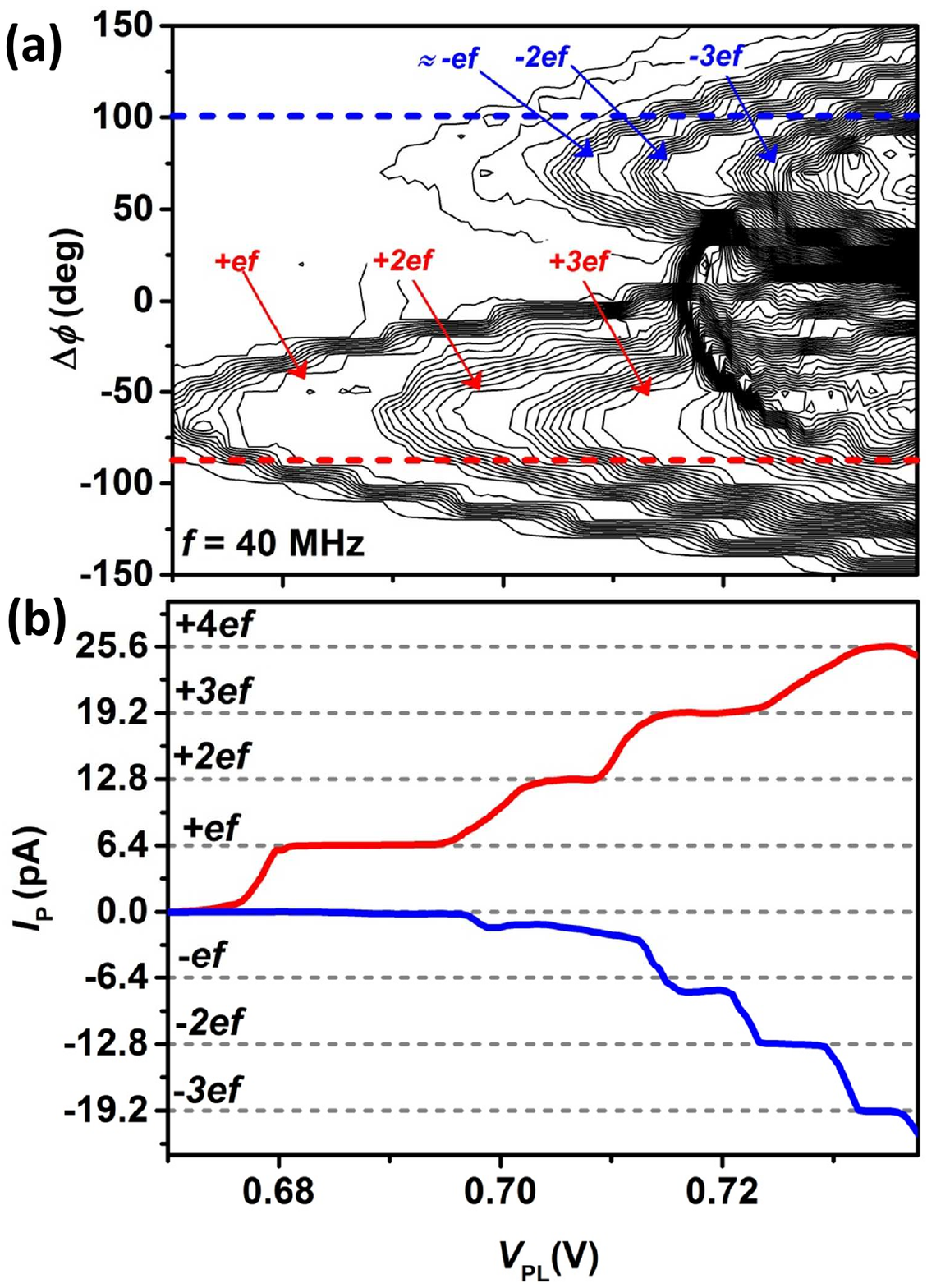}
\caption{(a) Pumped current obtained with the two-parameter drive at 40 MHz as a function of $V_\textup{PL}$ and  $\Delta \phi$ for $\widetilde{V}_\textup{PL}/\widetilde{V}_\textup{BL}=0.71$, $V_\textup{C1}=0$~V and $V_\textup{C2}=0.35$~V. Contour lines are in steps of 500 fA. (b) Pumped current as a function of $V_\textup{PL}$ for $\Delta \phi=100$~deg (in blue) and $\Delta \phi=-86$~deg (in red). Traces taken from (a) as indicated by dashed lines. }
\label{fgr:phase}
\end{figure*}

\begin{figure*}[t]
\includegraphics[scale=0.45]{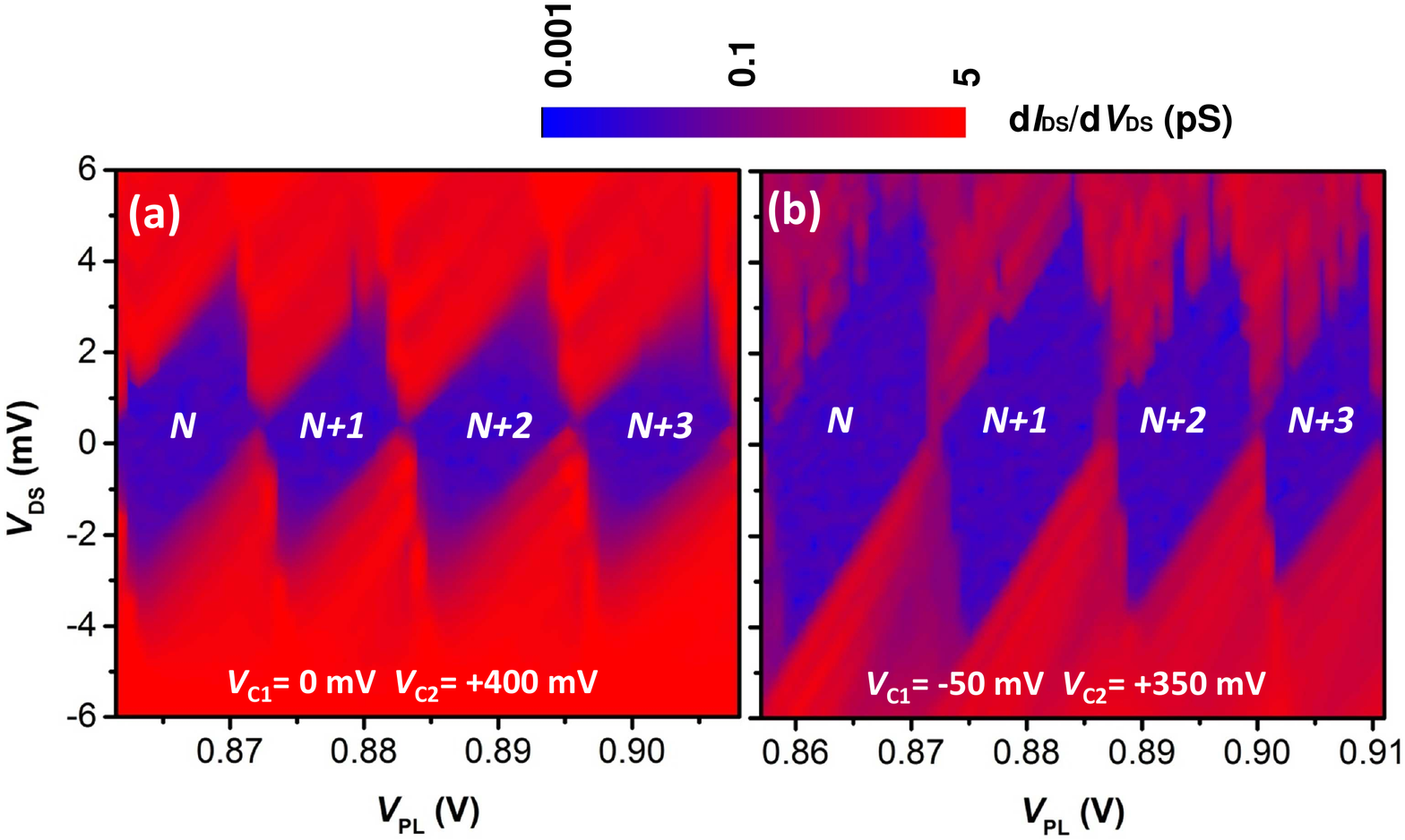}
\caption{Differential conductance of the quantum dot in the multi-electron regime as a function of the plunger gate voltage and the drain--source bias for (a) $V_\textup{C1}=0$, $V_\textup{C2}=400$~mV, (b) $V_\textup{C1}=-50$~mV, $V_\textup{C2}=350$~mV. For both datasets $V_\textup{BL}=820$~mV, $V_\textup{BR}=816$~mV. No driving signals are applied to the gates.}
\label{fgr:diam}
\end{figure*}
As discussed in the main article, by using two driving voltages we can obtain quantized current in either direction. The potential landscape experienced by the electrons in this configuration can be modelled by
\begin{equation}
\begin{array}{l}
\displaystyle \varphi_\textup{BL}(t)\propto\alpha_\textup{BL/BL}[V_\textup{BL}+\widetilde{V}_\textup{BL} \sin(2\pi f t)]+\alpha_\textup{BL/PL}[V_\textup{PL}+\widetilde{V}_\textup{PL} \sin  (2\pi f t+\Delta \phi)]+\alpha_\textup{BL/BR}V_\textup{BR}   \\
\displaystyle \varphi_\textup{PL}(t)\propto\alpha_\textup{PL/BL}[V_\textup{BL}+\widetilde{V}_\textup{BL} \sin(2\pi f t)]+\alpha_\textup{PL/PL}[V_\textup{PL}+\widetilde{V}_\textup{PL} \sin  (2\pi f t+\Delta \phi)]+\alpha_\textup{PL/BR}V_\textup{BR} \\
\displaystyle \varphi_\textup{BR}(t)\propto\alpha_\textup{BR/BL}[V_\textup{BL}+\widetilde{V}_\textup{BL} \sin(2\pi f t)]+\alpha_\textup{BR/PL}[V_\textup{PL}+\widetilde{V}_\textup{PL} \sin  (2\pi f t+\Delta \phi)]+\alpha_\textup{BR/BR}V_\textup{BR}
\end{array} 
\label{eq:potentials}
\end{equation}
where the $\alpha$ factors are constants representing the couplings between gates and electrons at different positions. Thus, the effective phase differences between the individual potentials are functions of $\Delta \phi$, $\widetilde{V}_\textup{PL}$ and $\widetilde{V}_\textup{BL}$. Hence, the direction of the electron transfer can be experimentally controlled by tuning these variables, independently of the drain--source bias that can be conveniently left at zero. In the measurements shown in the main article, we have $\widetilde{V}_\textup{BL}\gg\widetilde{V}_\textup{PL}$, so that the transfer occurs from source to drain irrespective of the choice of $\Delta \phi$. This happens because the left barrier dynamics dominate and define a pumping protocol similar to the single sinusoidal drive of Fig. 1(c) and (e) of the main article. By contrast, in Fig.~\ref{fgr:phase} we report data where bidirectional pumping takes place upon modification of $\Delta \phi$. For these measurements $\widetilde{V}_\textup{PL}\approx\widetilde{V}_\textup{BL}$ and the way the two waveforms combine is primarily dictated by their phases. We note that the quantization at negative currents does not appear as good as for the positive direction. We attribute this behaviour to the fact that the input barrier cannot be controlled as efficiently for the negative current as for the positive, since no drive signal is directly applied to BR.
\subsection{Effect of confinement bias on charging energy}

\begin{figure*}[t]
\includegraphics[scale=0.60]{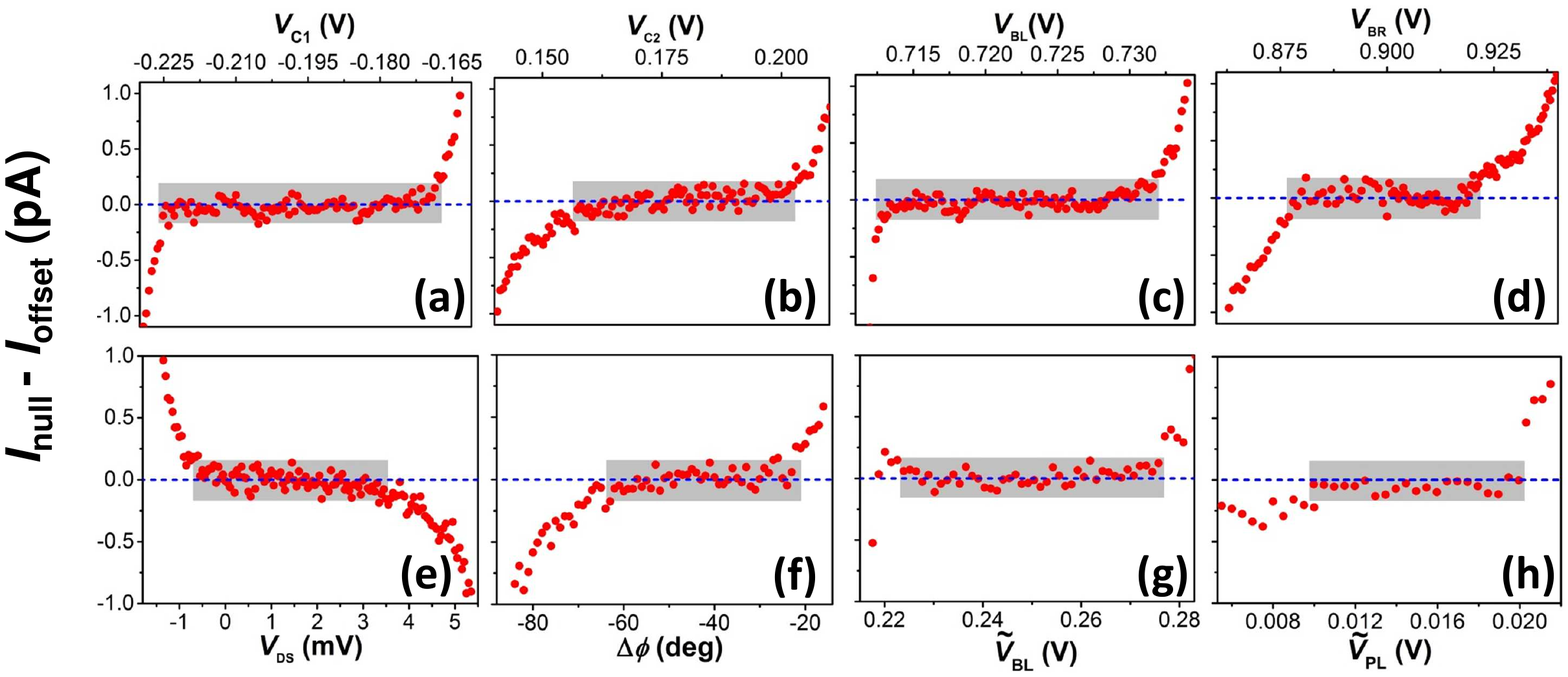}
\caption{Coarse null measurements of the pumped current at 0.5 GHz as a function of (a) $V_\textup{C1}$, (b) $V_\textup{C2}$, (c) $V_\textup{BL}$, (d) $V_\textup{BR}$, (e) $V_\textup{DS}$, (f) $\Delta\phi$, (g) $\tilde{V}_\textup{BL}$, and (h) $\tilde{V}_\textup{PL}$. A fixed offset current $I_\textup{offset}=550$~fA is subtracted from each trace instead of taking readings for multiple ON-OFF cycles. The integration time for an individual data point is 4 s. Dashed lines are guides for the eye to highlight the zero current level. The grey shaded areas indicate the parameter ranges for which the deviation of the current from the expected value falls within $\pm 160$~fA.}
\label{fgr:stab}
\end{figure*}
As we discuss in the main article, gates  C1 and C2 are utilized to control the planar confinement of the quantum dot. A convenient way of demonstrating the effectiveness of these gates is the evaluation of the dot charging energy, $E_\textup{C}$, for different values of $V_\textup{C1}$ and $V_\textup{C2}$. In our MOS structure, the vertical extension of the electron gas induced at the Si/SiO$_2$ interface is very little affected by the bias, and typically these regions are modelled as two-dimensional electron gases~\cite{florisrmp}. Therefore, any variation in the total capacitance of the quantum dot, $C_\textup{$\Sigma$}$, has to occur via a modification of the planar extension of the dot itself. This results in a change in the charging energy according to the relation $E_\textup{C}=e^2/C_\textup{$\Sigma$}$. In Fig.~\ref{fgr:diam}(a) and (b), we compare the Coulomb diamonds of the dot in the multi-electron regime for two different confinement configurations. We observe that, for a fixed number of electrons in the dot ($N$), $E_\textup{C}$ increases from $\approx 3.2$~meV to $\approx 5.0$~meV  with decreasing $V_\textup{C1}$ and $V_\textup{C2}$. This ultimately demonstrates that gates C1 and C2 can be used to control the size of the dot.

\subsection{Quantization robustness}

Any system that is aimed at quantum metrological applications, should be able to provide a stable and quantized output for a comfortably large range of all the adjustable parameters. In our experiments, we work with a multidimensional parameter space that needs to be iteratively scanned to yield the best current quantization. When one considers that in our experiments the drain--source bias is typically set at zero and the reservoir gates, $V_\textup{SL}$ and $V_\textup{DL}$, are kept at fixed positive bias, the number of variables can still vary from 6 for the single-parameter drive (namely, $V_\textup{C1}$, $V_\textup{C2}$, $V_\textup{PL}$, $V_\textup{BL}$, $V_\textup{BR}$, $\tilde{V}_\textup{BL}$ ) to 8 for the double-parameter drive (where  $\Delta\phi$ and $\tilde{V}_\textup{PL}$ come into play). Although a detailed study of the dependence of the pump accuracy on these parameters is beyond the scope of this work, in Fig.~\ref{fgr:stab} we report the results of a preliminary study. The measurement results shown are obtained with the same null detection configuration as discussed in the main article but with much faster integration time ($\approx 4$~s per data point). Moreover, they are taken as individual sweeps of the parameter of interest, rather than as the mean of multiple readings over subsequent ON-OFF cycles. Despite these limitations, we find that each experimental variable produces a current plateau in a sufficiently large range to allow simultaneous adjustments of other parameters. Due to the modest averaging time in these measurements, the dominant component of the uncertainty is of random type.
%\bibliography{pump}

\newpage
\begin{thebibliography}{60}

\bibitem{max}
Maxwell,~J.~C. In \emph{The scientific papers of {J}ames {C}lerk {M}axwell};
  Niven,~W., Ed.; Cambridge University Press, 1890; Vol.~2; p 225
  %\relax
%\mciteBstWouldAddEndPuncttrue
%\mciteSetBstMidEndSepPunct{\mcitedefaultmidpunct}
%{\mcitedefaultendpunct}{\mcitedefaultseppunct}\relax
%\EndOfBibitem
\bibitem[Josephson(1962)]{josephson}
Josephson,~B.~D. \emph{Phys. Lett.} \textbf{1962}, \emph{1}, 251\relax
\bibitem[Klitzing{et~al.}(1980)Klitzing, Dorda, and Pepper]{hall}
Klitzing,~K.~v.; Dorda,~G.; Pepper,~M. \emph{Phys. Rev. Lett.} \textbf{1980},
  \emph{45}, 494--497
\bibitem[Zimmerman and Keller(2003)Zimmerman, and Keller]{e_metro}
Zimmerman,~N.~M.; Keller,~M.~W. \emph{Meas. Sci. Technol.} \textbf{2003},
  \emph{14}, 1237
\bibitem[Pekola{et~al.}(2013)Pekola, Saira, Maisi, Kemppinen,
  M\"ott\"onen, Pashkin, and Averin]{peko}
Pekola,~J.~P.; Saira,~O.-P.; Maisi,~V.~F.; Kemppinen,~A.; M\"ott\"onen,~M.;
  Pashkin,~Y.~A.; Averin,~D.~V. \emph{Rev. Mod. Phys.} \textbf{2013},
  \emph{85}, 1421--1472
\bibitem[Devoret{et~al.}(1992)Devoret, Esteve, and Urbina]{devo}
Devoret,~M.~H.; Esteve,~D.; Urbina,~C. \emph{Nature} \textbf{1992}, \emph{360},
  547
\bibitem[Keller{et~al.}(1996)Keller, Martinis, Zimmerman, and
  Steinbach]{zimmer}
Keller,~M.~W.; Martinis,~J.~M.; Zimmerman,~N.~M.; Steinbach,~A.~H. \emph{Appl.
  Phys. Lett.} \textbf{1996}, \emph{69}, 1804--1806
\bibitem[Pekola{et~al.}(2008)Pekola, Vartiainen, M{\"{o}}tt{\"{o}}nen,
  Saira, Meschke, and Averin]{pek}
Pekola,~J.~P.; Vartiainen,~J.~J.; M{\"{o}}tt{\"{o}}nen,~M.; Saira,~O.-P.;
  Meschke,~M.; Averin,~D.~V. \emph{Nat. Phys.} \textbf{2008}, \emph{4},
  120
\bibitem[Maisi (2009)Maisi, Pashkin, Kafanov, Tsai, and
  Pekola]{ville}
Maisi,~V.~F.; Pashkin,~Y.~A.; Kafanov,~S.; Tsai,~J.-S.; Pekola,~J.~P. \emph{New
  J. Phys.} \textbf{2009}, \emph{11}, 113057
\bibitem[Connolly (2013)Connolly, Chiu, Giblin, Kataoka,
  Fletcher, Chua, Griffiths, Jones, Fal'ko, Smith, and Janssen]{graphene}
Connolly,~M.~R.; Chiu,~K.~L.; Giblin,~S.~P.; Kataoka,~M.; Fletcher,~J.~D.;
  Chua,~C.; Griffiths,~J.~P.; Jones,~G. A.~C.; Fal'ko,~V.~I.; Smith,~C.~G.;
  Janssen,~T. J. B.~M. \emph{Nat. Nano.} \textbf{2013}, \emph{8}, 417
\bibitem[Lansbergen{et~al.}(2012)Lansbergen, Ono, and Fujiwara]{lans}
Lansbergen,~G.~P.; Ono,~Y.; Fujiwara,~A. \emph{Nano Letters} \textbf{2012},
  \emph{12}, 763--768
\bibitem[Roche{et~al.}(2013)Roche, Riwar, Voisin, Dupont-Ferrier,
  Wacquez, Vinet, Sanquer, Splettstoesser, and Jehl]{roche}
Roche,~B.; Riwar,~R.-P.; Voisin,~B.; Dupont-Ferrier,~E.; Wacquez,~R.;
  Vinet,~M.; Sanquer,~M.; Splettstoesser,~J.; Jehl,~X. \emph{Nat. Comm.}
  \textbf{2013}, \emph{4}, 1581
\bibitem[Tettamanzi{et~al.}(2014)Tettamanzi, Wacquez, and Rogge]{giu}
Tettamanzi,~G.~C.; Wacquez,~R.; Rogge,~S. \emph{arXiv.org/cond-matt}
  \textbf{2014}, \emph{1401.3080}
\bibitem[Ono and Takahashi(2003)Ono, and Takahashi]{ono}
Ono,~Y.; Takahashi,~Y. \emph{Appl. Phys. Lett.} \textbf{2003}, \emph{82},
  1221--1223
\bibitem[Fujiwara{et~al.}(2004)Fujiwara, Zimmerman, Ono, and
  Takahashi]{fuji04}
Fujiwara,~A.; Zimmerman,~N.~M.; Ono,~Y.; Takahashi,~Y. \emph{Appl. Phys. Lett.}
  \textbf{2004}, \emph{84}, 1323--1325
\bibitem[Fujiwara{et~al.}(2008)Fujiwara, Nishiguchi, and Ono]{fuji08}
Fujiwara,~A.; Nishiguchi,~K.; Ono,~Y. \emph{Appl. Phys. Lett.} \textbf{2008},
  \emph{92}, 042102
\bibitem[Chan{et~al.}(2011)Chan, M{\"{o}}tt{\"{o}}nen, Kemppinen, Lai,
  Tan, Lim, and Dzurak]{kokwai}
Chan,~K.~W.; M{\"{o}}tt{\"{o}}nen,~M.; Kemppinen,~A.; Lai,~N.~S.; Tan,~K.~Y.;
  Lim,~W.~H.; Dzurak,~A.~S. \emph{Appl. Phys. Lett.} \textbf{2011}, \emph{98},
  212103
\bibitem[Jehl{et~al.}(2013)Jehl, Voisin, Charron, Clapera, Ray, Roche,
  Sanquer, Djordjevic, Devoille, Wacquez, and Vinet]{xavier}
Jehl,~X.; Voisin,~B.; Charron,~T.; Clapera,~P.; Ray,~S.; Roche,~B.;
  Sanquer,~M.; Djordjevic,~S.; Devoille,~L.; Wacquez,~R.; Vinet,~M. \emph{Phys.
  Rev. X} \textbf{2013}, \emph{3}, 021012
\bibitem[Kouwenhoven{et~al.}(1991)Kouwenhoven, Johnson, van~der Vaart,
  Harmans, and Foxon]{kou}
Kouwenhoven,~L.~P.; Johnson,~A.~T.; van~der Vaart,~N.~C.; Harmans,~C. J. P.~M.;
  Foxon,~C.~T. \emph{Phys. Rev. Lett.} \textbf{1991}, \emph{67},
  1626--1629
\bibitem[Shilton{et~al.}(1996)Shilton, Talyanskii, Pepper, Ritchie,
  Frost, Ford, Smith, and Jones]{saw}
Shilton,~J.~M.; Talyanskii,~V.~I.; Pepper,~M.; Ritchie,~D.~A.; Frost,~J. E.~F.;
  Ford,~C. J.~B.; Smith,~C.~G.; Jones,~G. A.~C. \emph{J. Phys.} \textbf{1996},
  \emph{8}, L531
\bibitem[Blumenthal{et~al.}(2007)Blumenthal, Kaestner, Li, Giblin,
  Janssen, Pepper, Anderson, Jones, and Ritchie]{blume}
Blumenthal,~M.~D.; Kaestner,~B.; Li,~L.; Giblin,~S.~P.; Janssen,~T. J. B.~M.;
  Pepper,~M.; Anderson,~D.; Jones,~G.; Ritchie,~D.~A. \emph{Nat. Phys.}
  \textbf{2007}, \emph{3}, 343
\bibitem[Kaestner{et~al.}(2008)Kaestner, Kashcheyevs, Amakawa,
  Blumenthal, Li, Janssen, Hein, Pierz, Weimann, Siegner, and
  Schumacher]{PTB_PRB}
Kaestner,~B.; Kashcheyevs,~V.; Amakawa,~S.; Blumenthal,~M.~D.; Li,~L.;
  Janssen,~T. J. B.~M.; Hein,~G.; Pierz,~K.; Weimann,~T.; Siegner,~U.;
  Schumacher,~H.~W. \emph{Phys. Rev. B} \textbf{2008}, \emph{77}, 153301
\bibitem[Wright{et~al.}(2008)Wright, Blumenthal, Gumbs, Thorn, Pepper,
  Janssen, Holmes, Anderson, Jones, Nicoll, and Ritchie]{samPRB}
Wright,~S.~J.; Blumenthal,~M.~D.; Gumbs,~G.; Thorn,~A.~L.; Pepper,~M.;
  Janssen,~T. J. B.~M.; Holmes,~S.~N.; Anderson,~D.; Jones,~G. A.~C.;
  Nicoll,~C.~A.; Ritchie,~D.~A. \emph{Phys. Rev. B} \textbf{2008}, \emph{78},
  233311
\bibitem[Kaestner{et~al.}(2009)Kaestner, Leicht, Kashcheyevs, Pierz,
  Siegner, and Schumacher]{PTB_apl}
Kaestner,~B.; Leicht,~C.; Kashcheyevs,~V.; Pierz,~K.; Siegner,~U.;
  Schumacher,~H.~W. \emph{Appl. Phys. Lett.} \textbf{2009}, \emph{94},
  012106
\bibitem[Wright{et~al.}(2009)Wright, Blumenthal, Pepper, Anderson,
  Jones, Nicoll, and Ritchie]{sam_par}
Wright,~S.~J.; Blumenthal,~M.~D.; Pepper,~M.; Anderson,~D.; Jones,~G. A.~C.;
  Nicoll,~C.~A.; Ritchie,~D.~A. \emph{Phys. Rev. B} \textbf{2009}, \emph{80},
  113303
\bibitem[Giblin{et~al.}(2010)Giblin, Wright, Fletcher, Kataoka, Pepper,
  Janssen, Ritchie, Nicoll, Anderson, and Jones]{giblin_njp}
Giblin,~S.~P.; Wright,~S.~J.; Fletcher,~J.~D.; Kataoka,~M.; Pepper,~M.;
  Janssen,~T. J. B.~M.; Ritchie,~D.~A.; Nicoll,~C.~A.; Anderson,~D.; Jones,~G.
  A.~C. \emph{New J. Phys.} \textbf{2010}, \emph{12}, 073013
  \bibitem[Giblin{et~al.}(2012)Giblin, Kataoka, Fletcher, See, Janssen,
  Griffiths, Jones, Farrer, and Ritchie]{giblin_natcom}
Giblin,~S.~P.; Kataoka,~M.; Fletcher,~J.~D.; See,~P.; Janssen,~T. J. B.~M.;
  Griffiths,~J.~P.; Jones,~G. A.~C.; Farrer,~I.; Ritchie,~D.~A. \emph{Nat.
  Comm.} \textbf{2012}, \emph{3}, 930
 \bibitem[Fletcher{et~al.}(2012)Fletcher, Kataoka, Giblin, Park, Sim,
  See, Ritchie, Griffiths, Jones, Beere, and Janssen]{npl_prb}
Fletcher,~J.~D.; Kataoka,~M.; Giblin,~S.~P.; Park,~S.; Sim,~H.-S.; See,~P.;
  Ritchie,~D.~A.; Griffiths,~J.~P.; Jones,~G. A.~C.; Beere,~H.~E.; Janssen,~T.
  J. B.~M. \emph{Phys. Rev. B} \textbf{2012}, \emph{86}, 155311
\bibitem[Kataoka{et~al.}(2011)Kataoka, Fletcher, See, Giblin, Janssen,
  Griffiths, Jones, Farrer, and Ritchie]{npl_prl}
Kataoka,~M.; Fletcher,~J.~D.; See,~P.; Giblin,~S.~P.; Janssen,~T. J. B.~M.;
  Griffiths,~J.~P.; Jones,~G. A.~C.; Farrer,~I.; Ritchie,~D.~A. \emph{Phys.
  Rev. Lett.} \textbf{2011}, \emph{106}, 126801
\bibitem[Angus{et~al.}(2007)Angus, Ferguson, Dzurak, and Clark]{angus}
Angus,~S.~J.; Ferguson,~A.~J.; Dzurak,~A.~S.; Clark,~R.~G. \emph{Nano Letters}
  \textbf{2007}, \emph{7}, 2051--2055
\bibitem[Lim{et~al.}(2009)Lim, Zwanenburg, Huebl, M{\"{o}}tt{\"{o}}nen,
  Chan, Morello, and Dzurak]{lim}
Lim,~W.~H.; Zwanenburg,~F.~A.; Huebl,~H.; M{\"{o}}tt{\"{o}}nen,~M.;
  Chan,~K.~W.; Morello,~A.; Dzurak,~A.~S. \emph{App. Phys. Lett.}
  \textbf{2009}, \emph{95}, 242102
\bibitem[Yang{et~al.}(2013)Yang, Rossi, Ruskov, Lai, Mohiyaddin, Lee,
  Tahan, Klimeck, Morello, and Dzurak]{our_natcom}
Yang,~C.~H.; Rossi,~A.; Ruskov,~R.; Lai,~N.~S.; Mohiyaddin,~F.~A.; Lee,~S.;
  Tahan,~C.; Klimeck,~G.; Morello,~A.; Dzurak,~A.~S. \emph{Nat. Comm.}
  \textbf{2013}, \emph{4}, 2069
\bibitem[Yamahata{et~al.}(2011)Yamahata, Nishiguchi, and Fujiwara]{yama}
Yamahata,~G.; Nishiguchi,~K.; Fujiwara,~A. \emph{Appl. Phys. Lett.}
  \textbf{2011}, \emph{98}, 222104
\bibitem[Fricke{et~al.}(2013)Fricke, Wulf, Kaestner, Hohls, Mirovsky,
  Mackrodt, Dolata, Weimann, Pierz, Siegner, and Schumacher]{fricke}
Fricke,~L.; Wulf,~M.; Kaestner,~B.; Hohls,~F.; Mirovsky,~P.; Mackrodt,~B.;
  Dolata,~R.; Weimann,~T.; Pierz,~K.; Siegner,~U.; Schumacher,~H.~W.
  \emph{arXiv.org/cond-matt} \textbf{2013}, \emph{1312.5669}
\bibitem[Kashcheyevs and Kaestner(2010)Kashcheyevs, and Kaestner]{casc}
Kashcheyevs,~V.; Kaestner,~B. \emph{Phys. Rev. Lett.} \textbf{2010},
  \emph{104}, 186805
\bibitem[Lewis and Riesenfeld(1969)Lewis, and Riesenfeld]{math}
Lewis,~H.~R.; Riesenfeld,~W.~B. \emph{J. Math. Phys.} \textbf{1969}, \emph{10},
  1458
\bibitem[Hollenberg{et~al.}(2006)Hollenberg, Greentree, Fowler, and
  Wellard]{holle}
Hollenberg,~L. C.~L.; Greentree,~A.~D.; Fowler,~A.~G.; Wellard,~C.~J.
  \emph{Phys. Rev. B} \textbf{2006}, \emph{74}, 045311
\bibitem[Pla{et~al.}(2012)Pla, Tan, Dehollain, Lim, Morton, Jamieson,
  Dzurak, and Morello]{pla}
Pla,~J.~J.; Tan,~K.~Y.; Dehollain,~J.~P.; Lim,~W.~H.; Morton,~J. J.~L.;
  Jamieson,~D.~N.; Dzurak,~A.~S.; Morello,~A. \emph{Nature} \textbf{2012},
  \emph{489}, 541
\bibitem[Maune{et~al.}(2012)Maune, Borselli, Huang, Ladd, Deelman,
  Holabird, Kiselev, Alvarado-Rodriguez, Ross, Schmitz, Sokolich, Watson,
  Gyure, and Hunter]{maune}
Maune,~B.~M.; Borselli,~M.~G.; Huang,~B.; Ladd,~T.~D.; Deelman,~P.~W.;
  Holabird,~K.~S.; Kiselev,~A.~A.; Alvarado-Rodriguez,~I.; Ross,~R.~S.;
  Schmitz,~A.~E.; Sokolich,~M.; Watson,~C.~A.; Gyure,~M.~F.; Hunter,~A.~T.
  \emph{Nature} \textbf{2012}, \emph{481}, 344
  \bibitem{florisrmp}  Zwanenburg, F. A.; Dzurak, A. S.; Morello, A.; Simmons, M. Y.; Hollenberg, L. C. L.; Klimeck, G.; Rogge, S.; Coppersmith, S. N.; Eriksson, M. A. \textit{Rev. Mod. Phys.}~\textbf{2013}~\textit{85}, 961--1019.
\end{thebibliography}

\end{document}